\documentclass[final]{ws-ijmpc}
\usepackage[english]{babel}
\usepackage[utf8]{inputenc}
\usepackage{graphicx}
\usepackage{color}
\usepackage[super,compress]{cite}
\usepackage[breaklinks]{hyperref}
\hypersetup{colorlinks,urlcolor=black,citecolor=black,linkcolor=black,filecolor=black}
\usepackage{breakurl}
\addto\captionsenglish{\renewcommand{\bibname}{References}}

\newcommand{\stp}{\text{STP}}
\newcommand{\tl}{\text{TL}}

\newcommand{\go}{\text{GO}}
\newcommand{\cif}{\text{CAR IN FRONT}}
\newcommand{\stopp}{\text{STOP}}
\newcommand{\aggr}{\textit{aggressive}}
\newcommand{\care}{\textit{careful}}
\newcommand{\BW}{B_\text{w}}
\newcommand{\SW}{S_\text{w}}

\begin{document}
\markboth{Enrique Pazos}
{Influence of driver behavior in the emergence of traffic gridlocks}

\catchline{}{}{}{}{}

\title{Influence of driver behavior in the emergence of traffic gridlocks}

\author{Enrique Pazos}

\address{Instituto de Investigación en Ciencias Físicas y Matemáticas, Universidad de San Carlos de Guatemala, Edificio T-1, Ciudad Universitaria Z. 12\\
Guatemala, Guatemala 01012\\
epazos@ecfm.usac.edu.gt}

\maketitle

\begin{history}
\received{Day Month Year}
\revised{Day Month Year}
\end{history}

\begin{abstract}
We present a microscopic driving algorithm that prescribes the acceleration using three parameters: the distance to the leading vehicle, to the next traffic light and to the nearest stopping point when the next traffic light is in the red phase. We apply this algorithm to construct decision trees that enable two driving behaviors: aggressive and careful. The focus of this study is to analyze the amount of aggressive drivers that are needed in order to generate a traffic gridlock in a portion of a city with signalized intersections. At rush hour, aggressive drivers will enter the intersection regardless if they have enough time or space to clear it. When their traffic light changes they block other drivers, thus providing the conditions for a gridlock to develop. We find that gridlocks emerge even with very few aggressive drivers present. These results support the idea of promoting good driving behavior to avoid heavy congestion during rush hours.

\end{abstract}

\ccode{PACS Nos.: 02.70.-c, 89.65.-a, 05.65.+b}

\section{Introduction}
Traffic congestion is a problem in all mayor cities of the world. It has been studied using theoretical and empirical approaches with a variety of methods that include statistical physics, numerical analysis, fluid dynamics and cellular automata, to mention a few. A comprehensive review has been done in Refs.~\refcite{chowdhury2000statistical,helbing2001traffic}.

One of the most widely used methods to study traffic dynamics is cellular automata, from the pioneering work of Nagel and Schreckenberg~\cite{nagel1992cellular} to many subsequent improvements and extensions, see Refs.~\refcite{biham1992self,esser1997microscopic,chowdhury1999self,mhirech2013vehicular} and references therein. Cellular automata have the advantage that traffic rules are stated in fairly simple formulae whose computer implementation can run very fast. Other approach to traffic simulations are microscopic models, where the acceleration is given at a particular time as a function of parameters such as distances and velocities of the neighboring vehicles. Some examples are the follow-the-leader models~\cite{pipes1967car}, the Newell model~\cite{newell2002simplified} and the intelligent driver model~\cite{treiber2000congested}. Microscopic formulations have the advantage of giving a fine-grained time evolution which is a more realistic description of the motion~\cite{ehlert2001microscopic}.

In this study we will use a microscopic model to analyze the dynamics of an urban traffic scenario with signalized intersections, in which streets follow a Cartesian lay out and are single lanes with alternating senses of motion. The focus of our analysis is the influence of driver behavior in the emergence of gridlocks. To this end we devised a driving algorithm that specifies the acceleration based on three parameters which are the following distances: to the leading vehicle, to the next traffic light and to the nearest stopping point when the next traffic light is in the red phase. The advantage and motivation of this approach is that it enables the formulation of two driving behaviors which we call~\cite{ehlert2001microscopic} \aggr{} and \care{}. The aggressive driver follows the leading vehicle at a safe distance but will never care about blocking the intersection. This is a problem at  rush hour congestion, specially when his traffic light turns red and he has not cleared the intersection. When this happens the aggressive driver not only is standing still but also does not allow cars in the perpendicular lane to advance, hence providing the conditions for a gridlock to develop. On the other hand, the careful driver will never block the intersection, she will wait until having enough distance headway so that she can effectively clear the intersection without ever blocking it.

We set up runs with different amounts of aggressive and careful drivers and check whether a gridlock develops. Since traffic is the result of many factors and interactions, we also take into account the effects of taking a turn and the traffic light scheme. Our main conclusion is that gridlocks emerge even when very few aggressive drivers are present, while there is none when all drivers are careful.

This papers is organized as follows: section~\ref{sec:algorithm} describes our proposed driving scheme, the integration of the equations of motion, the computation of the acceleration and the decision trees that enable the aggressive and careful behaviors. Section~\ref{sec:results} presents the results for a freeway test case and the conditions that give rise to gridlocks in city traffic. We conclude in section~\ref{sec:conclusion}.




\section{Driving algorithm}
\label{sec:algorithm}
\begin{figure}
\centering
\includegraphics[width=0.7\textwidth]{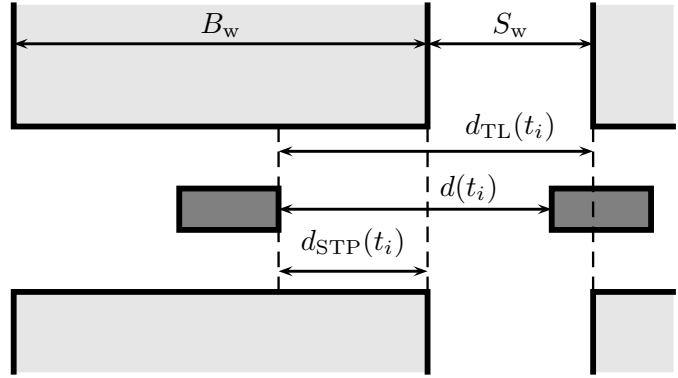}
\caption{Schematic representation of the intersection of two perpendicular streets. Cars are drawn as dark rectangles, the sense of motion is to the right. $\BW$ and $\SW$ are block and street width, respectively, $d(t_i)$ is the distance to the leading vehicle at the $i$-th iteration, $d_\text{TL}(t_i)$ is the distance to the traffic light and $d_\text{STP}(t_i)$ is the distance to the stopping location when the traffic light is red.}
\label{fig:streets}
\end{figure}

In order to account for the two different driving behaviors that we want to study, we devised a scheme to prescribe the acceleration of every vehicle based upon the three distances depicted in Fig.~\ref{fig:streets}. In freeway traffic the main parameter that determines the acceleration is the distance $d(t_i)$ to the leading vehicle\cite{helbing2001traffic}. In the model we propose here we add two more parameters, one is the distance $d_\tl(t_i)$ to the next traffic light and the other is the distance $d_\stp(t_i)$ to the stopping point when the traffic light is in the red phase. Notice that as long as the vehicle has not entered the intersection, these two distances satisfy $d_\tl(t_i)-d_\stp(t_i) = S_\text{w}$. However, when the car enters the intersection, $d_\stp(t_i)$ is reset accordingly for the next block, i.e. now the distances satisfy $d_\stp(t_i)-d_\tl(t_i) = B_\text{w}$. In other words, the difference between these distances is used to determine whether the vehicle has entered the intersection or not. This feature is used in the formulation of the driving behaviors (see Figs.~\ref{fig:aggr} and \ref{fig:care}).

Once the acceleration is known, the position and velocity of every vehicle is integrated using the Euler method, such that
\begin{eqnarray}
x(t_{i+1}) &=& x(t_i) + v(t_i) \, \delta t, \label{eq:integx} \\
v(t_{i+1}) &=& v(t_i) + a(t_i) \, \delta t, \label{eq:integv} \\
t_{i+1} &=& t_i + \delta t,
\end{eqnarray}
where $x(t_i)$, $v(t_i)$ and $a(t_i)$ are the position, velocity and acceleration for a given vehicle at the $i$-th iteration. Similarly, the elapsed time is $t_i$ and $\delta t$ is the time step. Since the acceleration values do not vary largely during the simulation, this method is good enough to integrate the equations of motion.

\subsection{Acceleration modes}
For each vehicle, the acceleration $a(t_i)$ is set by one of three acceleration modes, which we call \go, \cif{} and \stopp. These modes are a way to mimic how the driver controls the vehicle in an urban traffic scenario. The motivation behind the acceleration modes is the flexibility they provide in order to construct a modular decision tree that implements the driving behaviors we are interested in, namely the \care{} and the \aggr{}.

The acceleration in each mode is set as follows:
\begin{itemize}
\item \go{}: The vehicle moves with constant acceleration $a_\text{go}$ until it reaches a maximum speed $v_\text{max}$. It is given as
\begin{eqnarray}
a(t_i) &=& \left\{ \begin{array}{ll}
0, & v(t_i) \geq v_\text{max} \\
a_\text{go}, & v(t_i) < v_\text{max}. \\
\end{array} \right.
\end{eqnarray}

\item \cif{}: The vehicle moves at a speed such that it keeps a safe distance to the leading vehicle\cite{pipes1967car}. The safe distance is the necessary space such that the time elapsed from bumper to bumper as measured at a fixed location is $\delta t_\text{s}$. The acceleration is
\begin{equation}
a(t_i) = \frac{d(t_i) - d_\text{s}(t_i)}{\delta t \, \delta t_\text{s}},
\end{equation}
where $d(t_i)$ is the distance to the leading vehicle (bumper to bumper) at a time $t_i$ (see Fig.~\ref{fig:streets}) and $d_\text{s}(t_i) = v(t_i) \, \delta t_\text{s}$ is the safe separation between the cars such that at the current speed $v(t_i)$ the vehicle would take a time $\delta t_\text{s}$ to traverse. In this mode, the cars keep the same time separation, which will be a larger (shorter) length when going at a faster (slower) speed. In practice, when taking one iteration step for the velocity, Eq.~(\ref{eq:integv}) reduces to $v(t_{i+1}) = d(t_i)/\delta t_\text{s}$. In other words, the speed is set to the value that is needed to keep a time $\delta t_\text{s}$ to the leading vehicle.

\item \stopp{}: A car will stop when the traffic light is red. The acceleration in this case will be
\begin{equation}
a(t_i) = -\frac{v(t_i)^2}{2 d_\text{STP}(t_i)},
\end{equation}
where $d_\text{STP}(t_i)$ is the distance to the point where the car has to be at rest (see Fig.~\ref{fig:streets}). By taking this approach, the car breaks smoothly, reaching zero speed as $d_\text{STP}(t_i)$ approaches zero. In this mode we set $v(t_i) =0$ whenever $d_\text{STP}(t_i) < d_\text{min}$ or $d(t_i) < d_\text{min}$. The first condition ensures that a car stops before entering an intersection and the second one avoids collisions. The fixed quantity $d_\text{min}$ is the minimum separation distance between cars.
\end{itemize}

We will show in Sec.~\ref{sec:results} that the acceleration modes reproduce the fundamental diagram for freeway traffic shown in Refs.~\refcite{helbing2001traffic,daganzo2005variational,geroliminis2008existence}.

\subsection{Guided by car or traffic light}
The decision as to which acceleration mode is used at a given time is determined by the values of the distances mentioned above, i.e. $d(t_i)$, $d_\tl(t_i)$ and $d_\stp(t_i)$. We assume that the process of deciding which acceleration mode will be used starts by determining what is closer: either the leading vehicle or the traffic light at the nearest intersection. If the leading vehicle is closer, the driver will just follow it at a safe distance, regardless of the traffic light. On the contrary, if the traffic light is closer, then the driver will disregard the leading vehicle and will maneuver according to the color of the signal. In the first case we say that the driver is \textit{car-guided}, and in the second case he is \textit{TL-guided} (Traffic Light). This decision is made at the beginning of each iteration by comparing the distances $d(t_i)$ and $d_\tl(t_i)$ (see Fig.~\ref{fig:streets})
\begin{equation}
\text{if } d(t_i) \left\{
\begin{array}{ll}
\leq d_\tl(t_i) \Rightarrow & \text{car-guided} \\
> d_\tl(t_i) \Rightarrow & \text{TL-guided}. 
\end{array} \right.
\end{equation}
Once this step is done the next ones will depend on the color of the traffic light. We consider red, yellow and green phases with duration $T_r$, $T_y$ and $T_g$, respectively. We use the traffic light color as the starting point of the decision trees that will implement the two types of driving behavior, i.e. careful and aggressive.

\subsection{Two types of driving behavior}
We focus our attention on the traffic flux and speed attained when individual drivers behave in one of two ways, which we call the aggressive and the careful types, following Ref.~\refcite{ehlert2001microscopic}. The distinction between them occurs when the driver has to make a decision at the same time that he is approaching the intersection of two perpendicular streets. We assume that the traffic flow at the intersection is controlled by a traffic light and that all streets have a single lane with a fixed sense of motion. When the vehicle density is high, there is a chance that some drivers will block the intersection, either moving slowly or standing still completely. If the potentially blocking driver, has enough green light time to clear the intersection, then he just goes through it at a low speed. The problem arises when the traffic light changes from green to yellow and red while the car is still traversing the intersection. At this point the vehicles waiting on the perpendicular lane have now the green light, but they will not be able to move because the intersection is blocked, thus increasing congestion.

We define the careful driving behavior or careful type as that in which the driver will never block the intersection. The careful driver achieves this by waiting to have enough distance headway to the leading vehicle so that she can effectively clear the intersection. The aggressive type is the opposite. This driver will follow the leading vehicle at a safe distance but he will not have any consideration about obstructing an intersection whatsoever.

From now on, for clarity, we will drop the time dependence from the distances considered.

\subsubsection{The aggressive type}
\begin{figure}
\begin{minipage}{0.5\textwidth}
\centering
car-guided
\fbox{\includegraphics[width=0.9\textwidth]{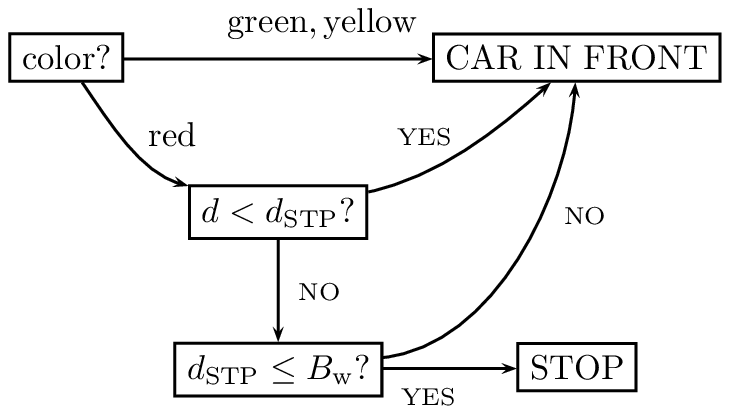}}
\end{minipage}%
\begin{minipage}{0.47\textwidth}
\centering
TL-guided
\fbox{\includegraphics[width=\textwidth]{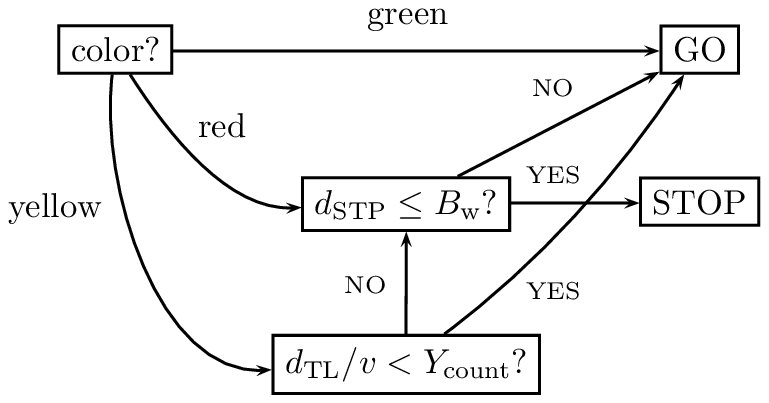}}
\end{minipage}
\caption{The aggressive behavior decision tree for a driver guided by the leading vehicle (left panel) and by traffic light (right panel).}
\label{fig:aggr}
\end{figure}

The decision trees for this driving conduct are shown in Fig.~\ref{fig:aggr}. In the car-guided tree the acceleration mode is basically \cif. The \stopp{} mode is reached when three conditions are met: the light is red, the leading vehicle has entered the intersection ($d>d_\stp$) and the driver has not entered it yet ($d_\stp \leq \BW$). Here, the aggressive conduct is manifest when the condition $d_\stp \leq \BW$ is false, which means that the driver is already going through the intersection and he will continue to move. The TL-guided tree is straightforward. The aggressive behavior is expressed when the light changes to red while the driver is already on the intersection $(d_\stp \leq \BW)$. The yellow light plays a role in this tree through the variable $Y_\text{count}$, which is the remaining time before the signal turns to red. The condition $d_\tl/v < Y_\text{count}$ tries to mimic the judging criterion that the driver uses to evaluate whether he has time to clear the intersection in the time that remains of yellow light.

\subsubsection{The careful type}
\begin{figure}
\begin{minipage}{0.45\textwidth}
\centering
car-guided
\fbox{\includegraphics[width=0.93\textwidth]{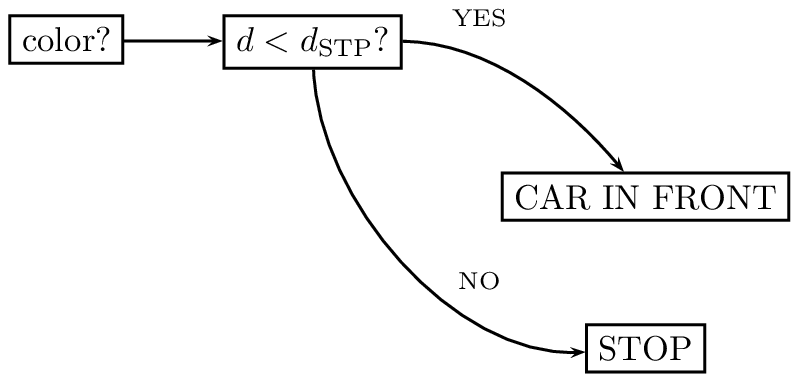}}
\end{minipage}%
\begin{minipage}{0.5\textwidth}
\centering
TL-guided
\fbox{\includegraphics[width=\textwidth]{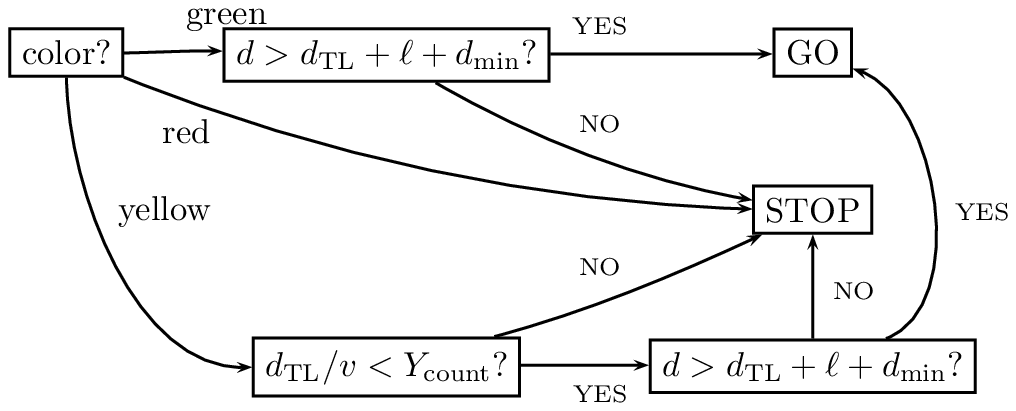}}
\end{minipage}
\caption{The careful behavior decision tree for a driver guided by the leading vehicle (left panel) and by traffic light (right panel).}
\label{fig:care}
\end{figure}

The decision trees for this behavior are illustrated in Fig.~\ref{fig:care}. The car-guided one is very simple: the driver either stops or follows the vehicle ahead regardless of the traffic light phase. The carefulness shows here in the fact that there is a single condition to go into \stopp{} mode, which is that the driver has not entered the intersection. This ensures that in the careful behavior the driver will always stop before going through the intersection. In the TL-guided panel, the driver executes more evaluations before crossing an intersection. If the signal is green she goes to \go{} mode only if she has enough space to accommodate herself after clearing the intersection completely $(d>d_\tl + \ell + d_{\min})$, where $\ell$ is the length of the car. We consider that a car blocks the intersection if any fraction of its length is over it. Red light means unconditional stop. In yellow phase the driver still evaluates whether she has enough time and space to safely clear the intersection altogether.

\section{Results and discussion}
\label{sec:results}
We implemented the above schemes using the C++ programming language. The setup consists of a grid-like city where all streets meet at perpendicular intersections in a Cartesian fashion. The extent of the city is 10 blocks in both $x$ and $y$ directions. Each street contains only a single lane, the sense of which alternates according to $(-1)^i\bf{\hat{x}}$ and $(-1)^j\bf{\hat{y}}$ for the $x$ and $y$ directions, respectively. The variables $i$ and $j$ are numbering integers in the interval $[0,9]$. All streets have periodic boundaries, which gives the city a 2-torus topology.

\subsection{Single lane with green lights}\label{sec:single}
In order to test our driving algorithm, we present first some results for a simplified setting. Using the city layout, let's consider the situation in which cars travel along a unique street only, and all traffic lights are set to green during the whole simulation. This setting mimics the conditions of a continuous road of length $L$ with periodic boundary conditions. We set $L=20(\BW+\SW)$ (see Fig.~\ref{fig:streets}), corresponding to a total length $L=2$~km, according to the values listed in Table~\ref{tab:params}.
\begin{table}
\tbl{Parameters used in the simulations. Their values are typical of different areas of Guatemala City.}
{\begin{tabular}{lp{0.4\columnwidth}l}
\toprule
parameter & & value \\ 
\colrule
$a_\text{go}$ & free acceleration & 1 m/s$^{2}$ \\ 
$\ell$ & car length & 5 m \\
$d_\text{min}$ & minimum distance between cars & 2 m \\
$\delta t_\text{s}$ & safe time interval between cars & 3 s \\
$\delta t$  & time step & 0.1 s \\
$v_\text{max}$ & maximum (free) speed & 11 m/s \\
$T_r$ & red light phase & 30 s\\
$T_y$ & yellow light phase & 5 s\\
$T_g$ & green light phase & 25 s\\
$\BW$ & block width & 90 m \\
$\SW$ & street width & 10 m \\
\botrule
\end{tabular}
\label{tab:params}}
\end{table}

We randomly distribute a number of vehicles $N$ with zero initial velocity along the street. This defines the density $\rho=N/L$. The cars start moving according to the rules outlined in Sec.~\ref{sec:algorithm}. We follow their motion for a three-hour simulated period, during which we compute the instantaneous average speed $\bar{v}(t_k)$ over all vehicles at intervals of 1~s according to
\begin{equation}
  \bar{v}(t_k) = \frac{1}{N} \sum_{j=1}^N v_j(t_k),
\end{equation}
where $v_j(t_k)$ is the speed of the $j$-th vehicle at a time $t_k$. Now we take the last 300~s of the simulation and compute the time averaged speed $V=\sum_k \bar{v}(t_k)/300$. These time averages have a standard deviation that is less than 4\% the value of $V$. Knowing $\rho$ and $V$ the traffic flow is given as $Q=\rho V$. By taking the last 300~s of a three-hour simulation we make sure that the transient part of the dynamics is over and we are left with the steady state.

\begin{figure}
\includegraphics[width=0.49\textwidth]{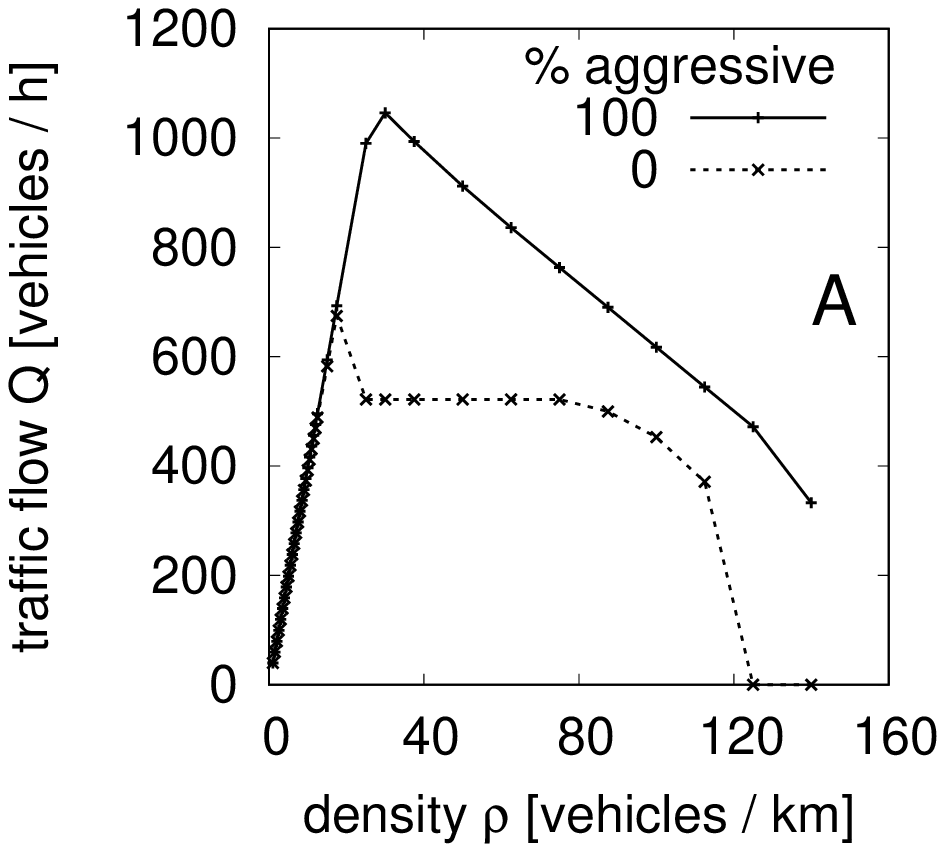}
\includegraphics[width=0.49\textwidth]{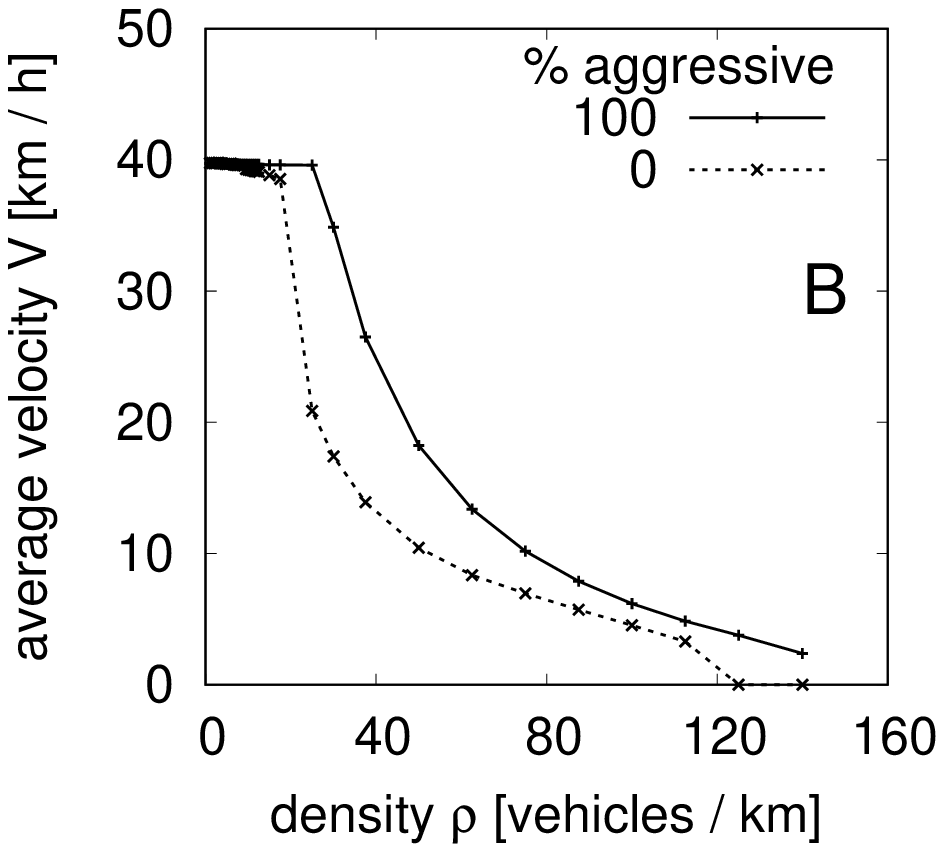}
\caption{Fundamental diagram for a single lane. Traffic flow $Q$ (panel A) and average velocity $V$ (panel B) vs. vehicle density $\rho$. The curves show the fundamental diagram for the two driving modes, namely, aggressive and careful.}
\label{fig:fluxSL}
\end{figure}

The results are shown in Fig.~\ref{fig:fluxSL}. We plot the traffic flow $Q$ (panel A) and the average velocity $V$ (panel B) versus the vehicle density $\rho$. We consider the two extreme cases: one where all drivers are aggressive and the other where all drivers are careful.  Our results for the aggressive type are in agreement with the fundamental diagrams in Refs.~\refcite{chowdhury1999self,chowdhury2000statistical,treiber2000congested,helbing2001traffic,mhirech2011traffic,qi2016impact}. The traffic flow $Q$ exhibits a linear growth at low densities, reaching a maximum when the safe separation is $d_\text{s} = v_\text{max}\, \delta t_\text{s}$. This corresponds to a critical density $\rho_\text{cr}=1/(\ell+d_\text{s}) \approx 11$~vehicles/km. From this point the flow decreases linearly until the density is $\rho_\text{max}=1/(\ell+d_\text{min}) \approx 143$~vehicles/km. When density reaches $\rho_\text{max}$ all vehicles are at their minimum separation $d_\text{min}$, thus they cannot be any closer and $Q$ goes to zero.

In the case where all drivers are careful (0\% aggressive), the traffic flow exhibits a different behavior. It starts increasing linearly at low densities, as in the previous case, but it drops to a constant value over a wide range of densities. This is due to the fact that careful drivers try to stop at every intersection if they don't have enough space to position themselves without blocking the intersection after crossing it. This behavior is desired at higher densities, thus it is not optimal for lower ones. Since we are interested in urban traffic and congestion at high densities we do not perform further optimization for the careful type at lower densities.

The fact that the aggressive type behavior reproduces the fundamental diagram for freeway traffic validates our ad-hoc driving algorithm formulation, which is qualitatively similar to those presented in Refs.~\refcite{helbing1998generalized,treiber2000congested}. Other feature we found in our simulations is the existence of stop-and-go waves, which are a well-known property of freeway traffic.~\cite{laval2010mechanism}

\subsection{City with turning vehicles and traffic lights}
We consider a square city of size $10 \times 10$ blocks, each block being a $\BW+\SW$ square per side. The total road length is $L=200(\BW+\SW)$, which amounts to 20~km. The density $\rho$ and average velocity $V$ are defined in the same way as in section~\ref{sec:single}.

Initially a number of vehicles $N$ are randomly distributed on all streets. All initial velocities are zero. We consider three driving behavior configurations for the $N$ drivers: all aggressive, half aggressive-half careful and all careful. We will name these configurations by the percentage of aggressive drivers they have, i.e. 100-, 50- and 0\%-aggressive, respectively. For each configuration we consider that there is a probability that a driver will take a turn at the next intersection. Ideally, the amount of turns a driver will take depends solely on the departure and destination points within the city. In reality it is influenced by other factors such as traffic incidents along the way, a preferred route or the driver's empirical knowledge of the state of the city traffic. To cover a wide range of turning probabilities we set up a series of runs, each with a fixed probability to take a turn. For instance, a run with 0\% turning probability means that all drivers will just go straight ahead, the amount of cars in each lane will remain constant throughout the simulated time. The probabilities considered here are 0, 10, 25, 50 and 75\%. For the type of city we are analyzing, i.e. streets consisting of a single lane with alternating and unique senses, a trip from one location to another can be accomplished by taking turns in 10 to 25\% of the traversed intersections.

Traffic lights are set up in two different ways: synchronized (sync-) and random (rand-) modes. In sync-mode all traffic lights change at the same time. Initially, the lights are in green phase for both senses along the $x$-direction, consequently they are in red phase for both senses along the $y$-direction. The duration of the three phases is the same for all traffic lights (see Table~\ref{tab:params}). In rand-mode each traffic light is out of phase by a random fraction of the cycle. The extent of the cycle is still the same for all traffic lights but now the shifting sequence has been randomly ordered.

\begin{figure}
\includegraphics[width=0.32\textwidth]{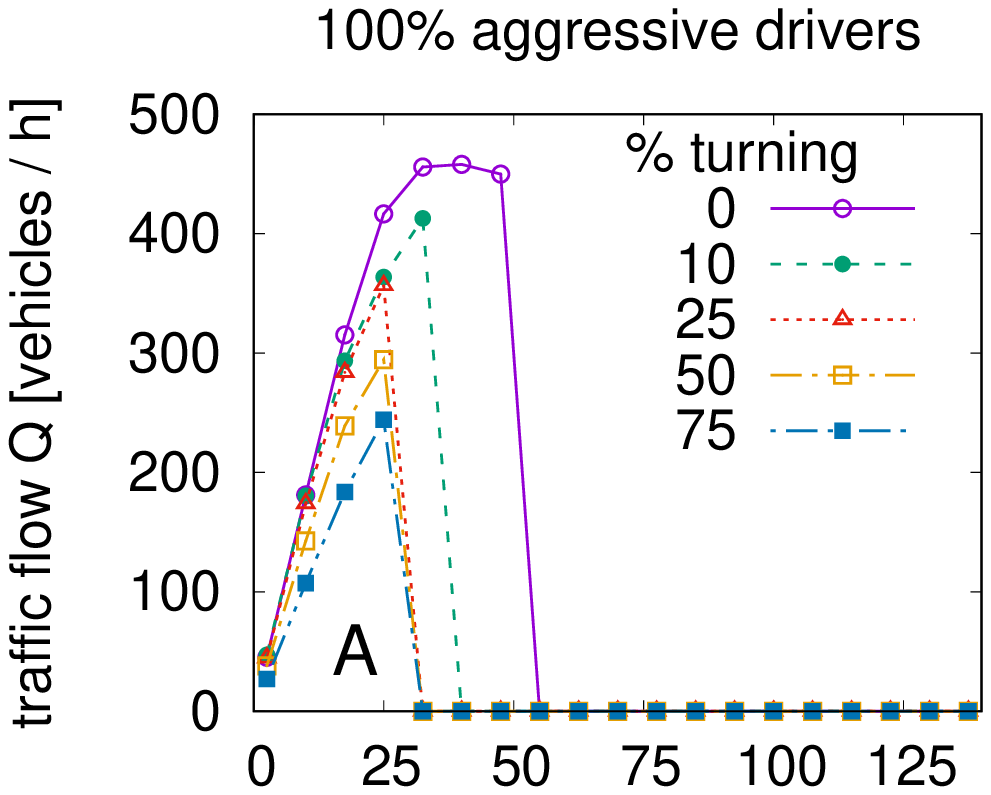}
\includegraphics[width=0.32\textwidth]{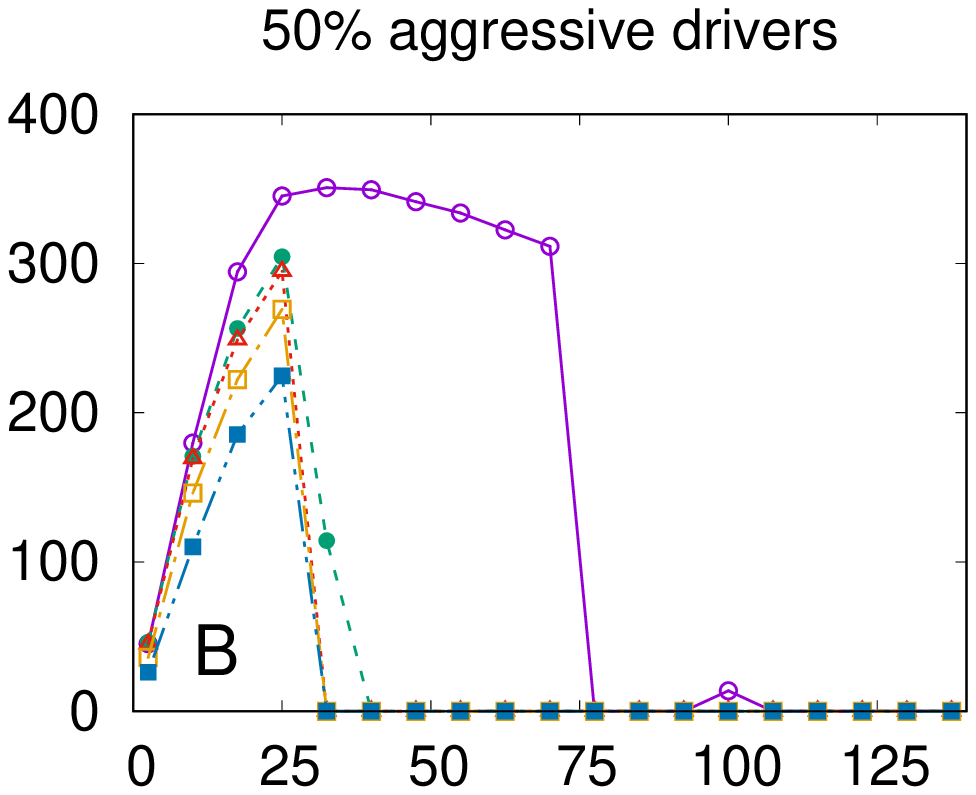}
\includegraphics[width=0.32\textwidth]{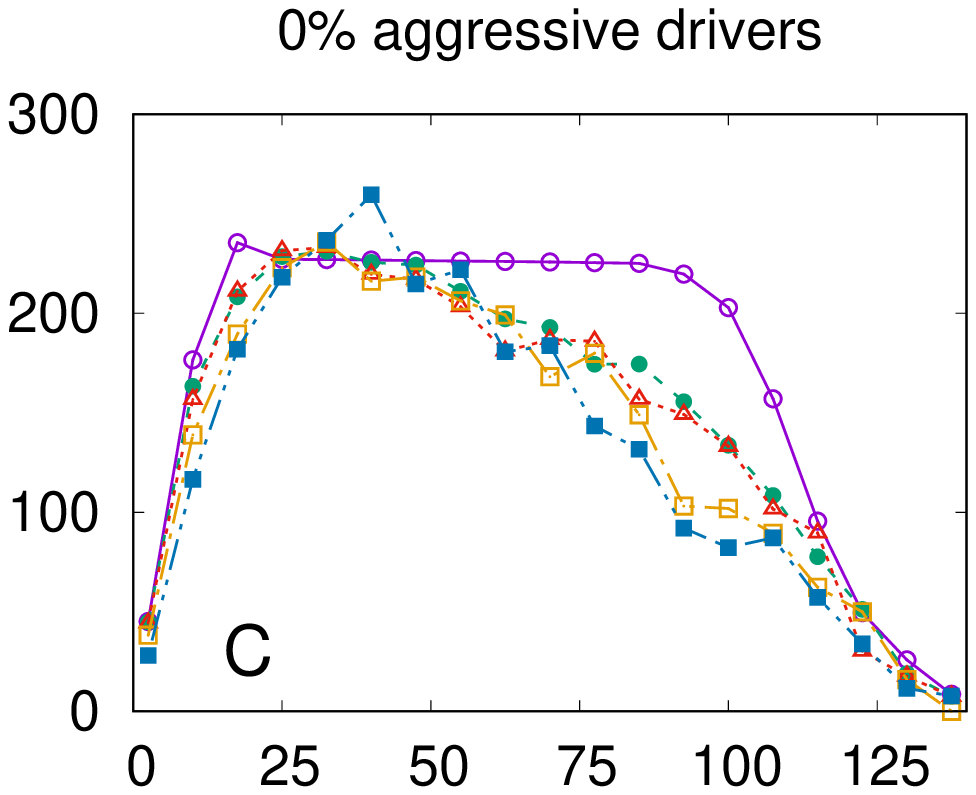}\\
\includegraphics[width=0.32\textwidth]{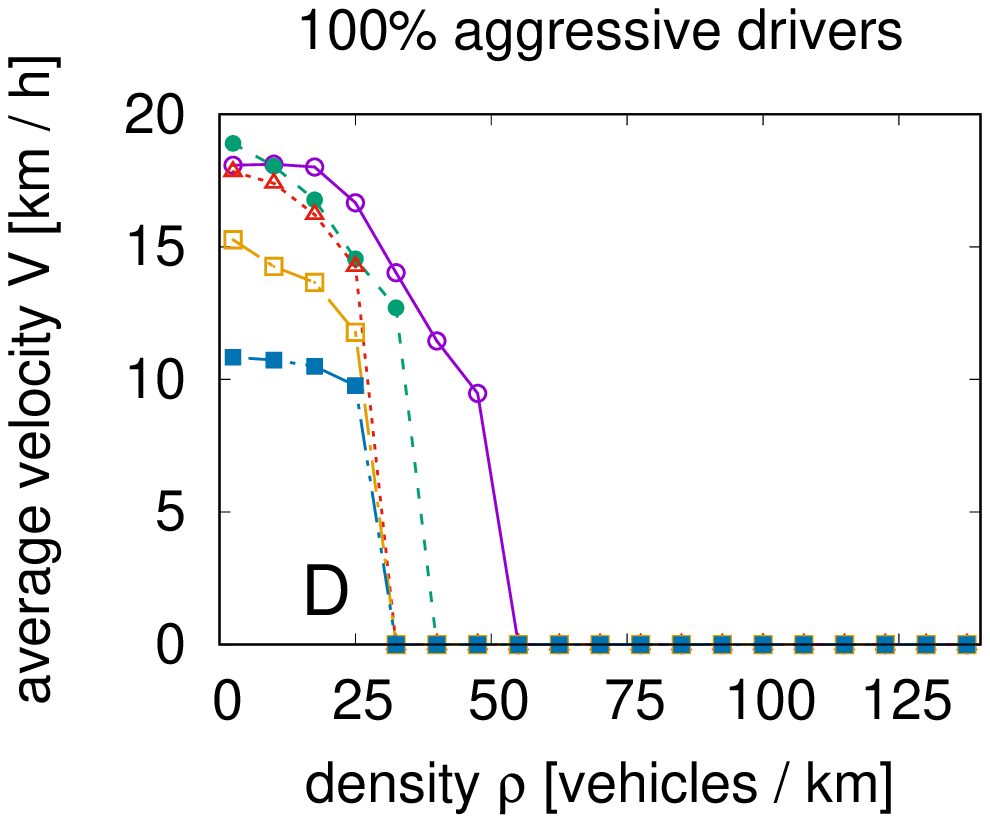}
\includegraphics[width=0.32\textwidth]{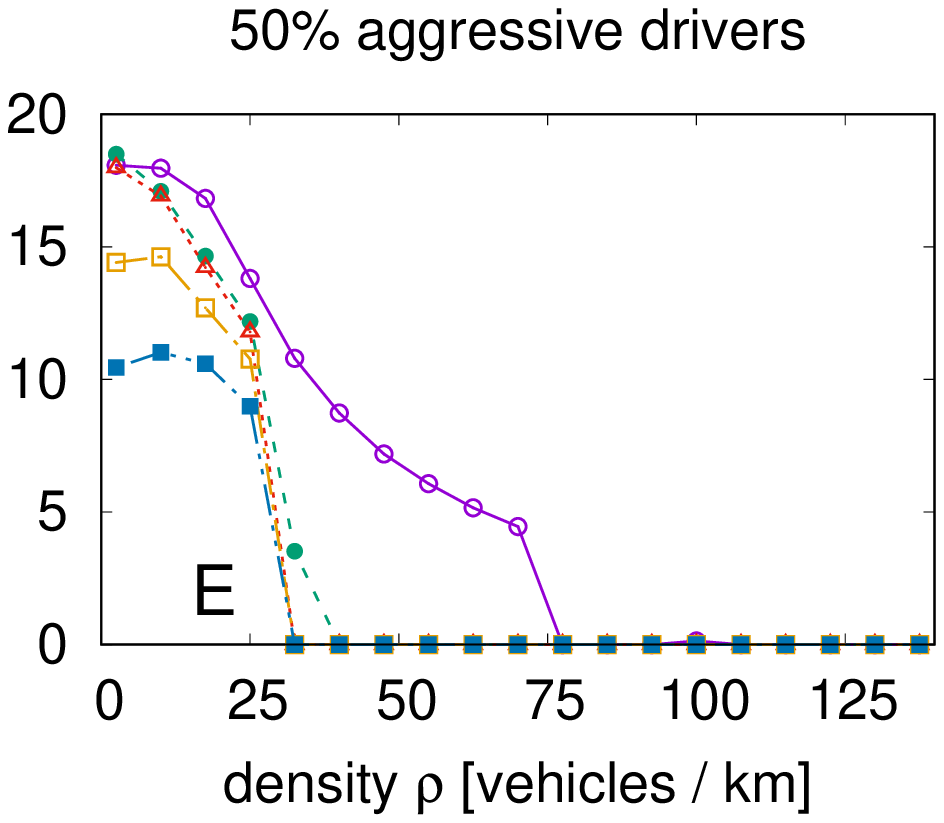}
\includegraphics[width=0.32\textwidth]{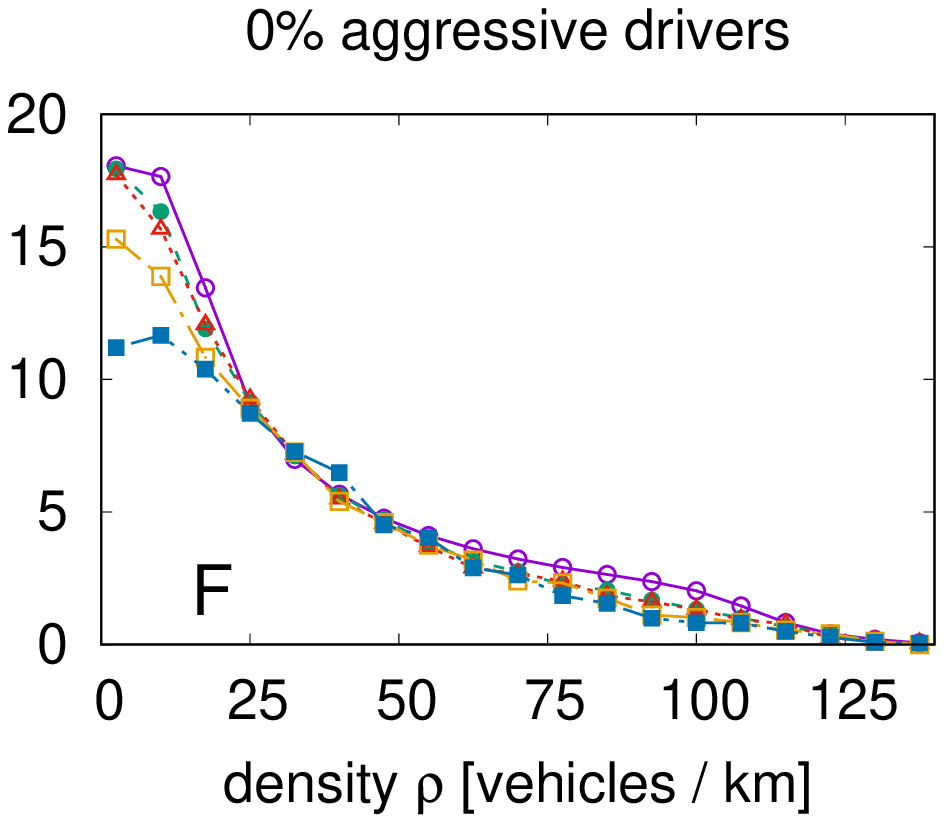}
\caption{Traffic flow (top row) and average velocity (bottom row) for the three configurations of aggressive and careful drivers using sync-mode traffic lights.}
\label{fig:sync}
\end{figure}

\begin{figure}
\includegraphics[width=0.32\textwidth]{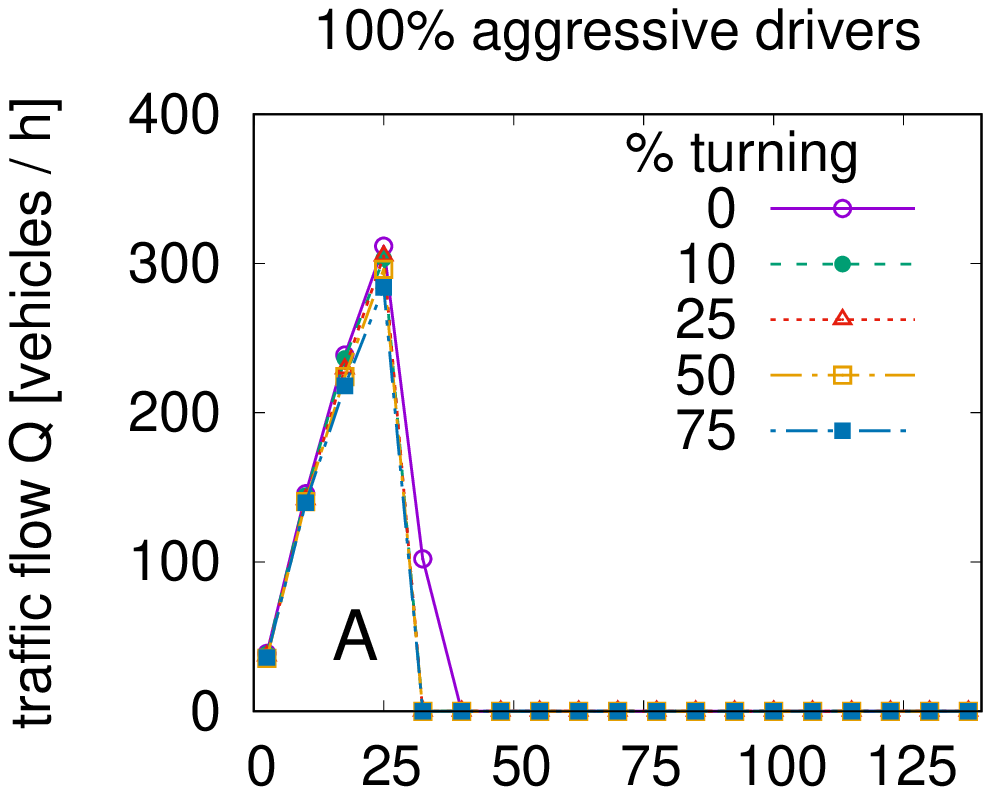}
\includegraphics[width=0.32\textwidth]{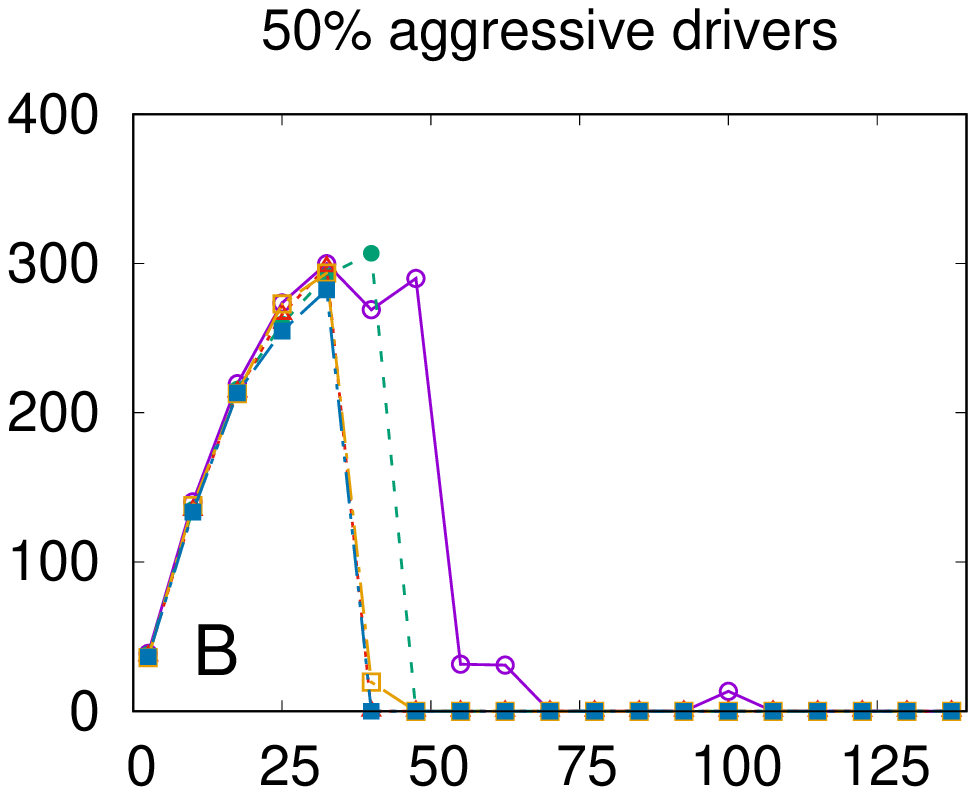}
\includegraphics[width=0.32\textwidth]{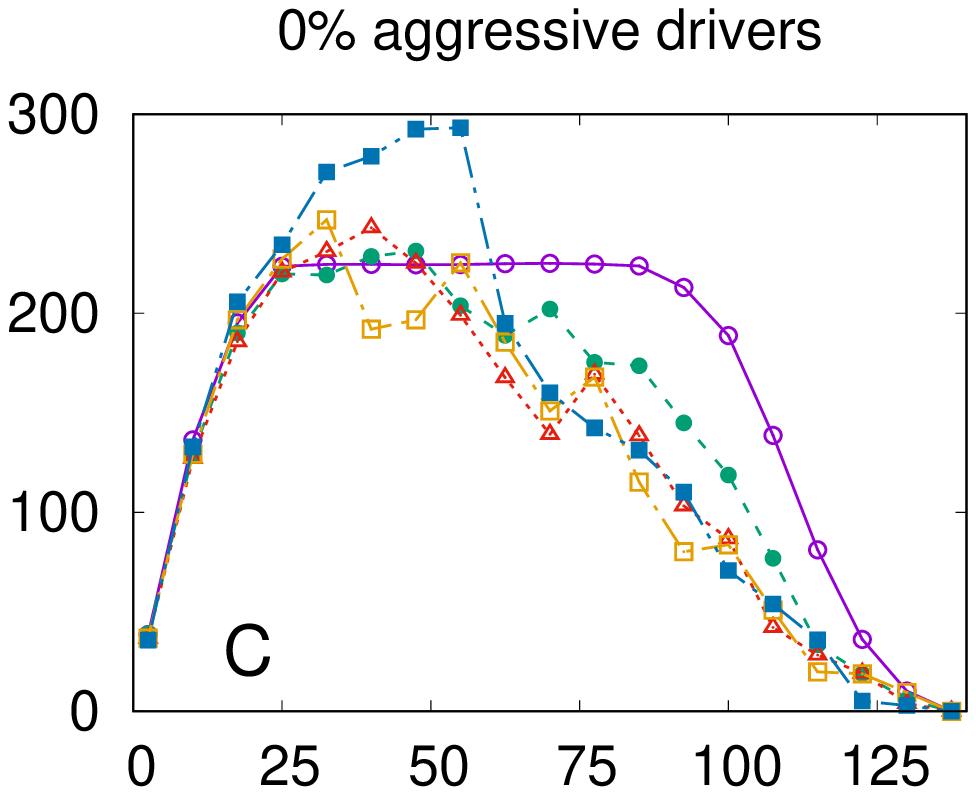}\\
\includegraphics[width=0.32\textwidth]{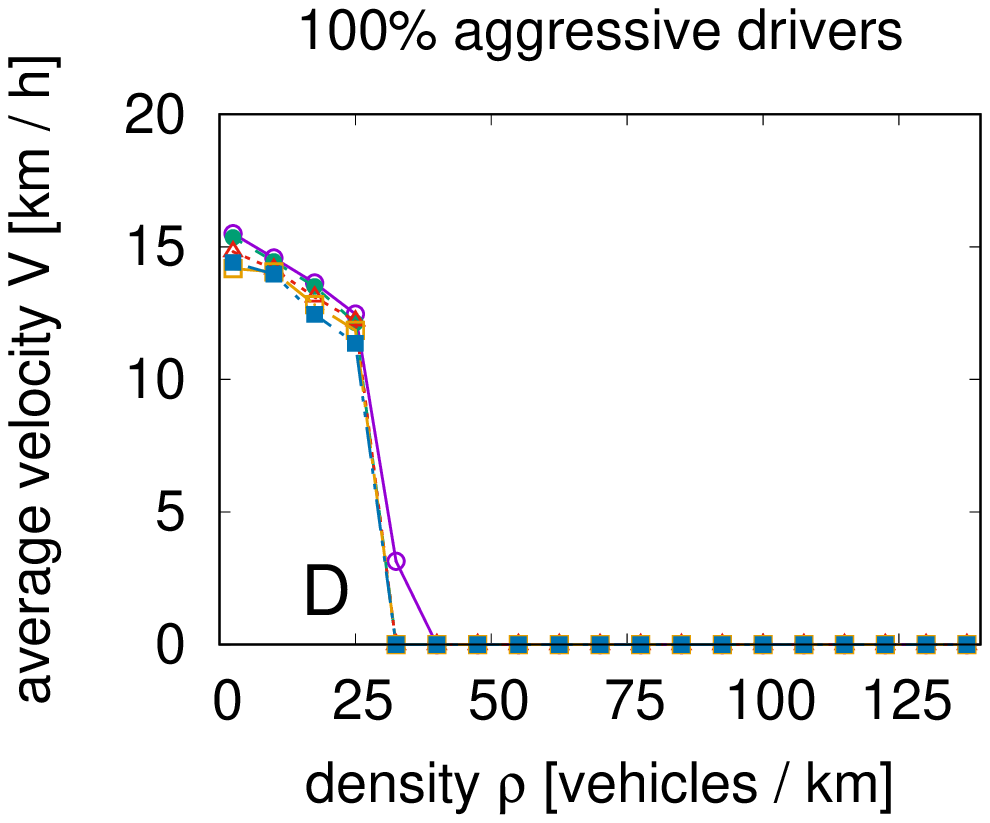}
\includegraphics[width=0.32\textwidth]{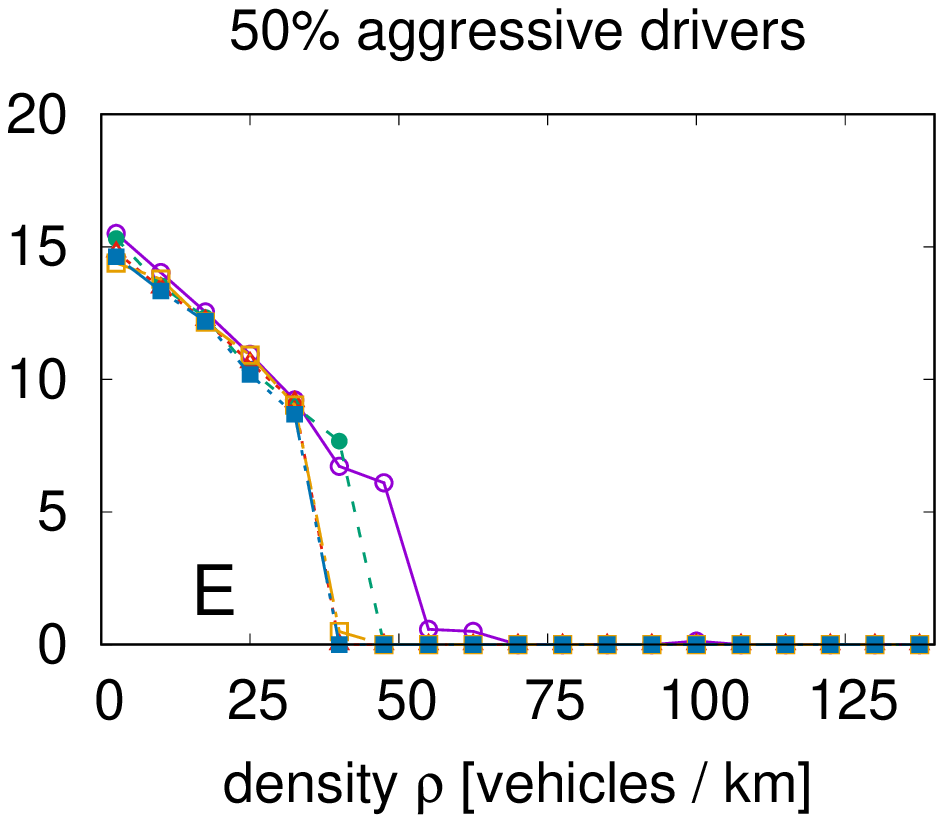}
\includegraphics[width=0.32\textwidth]{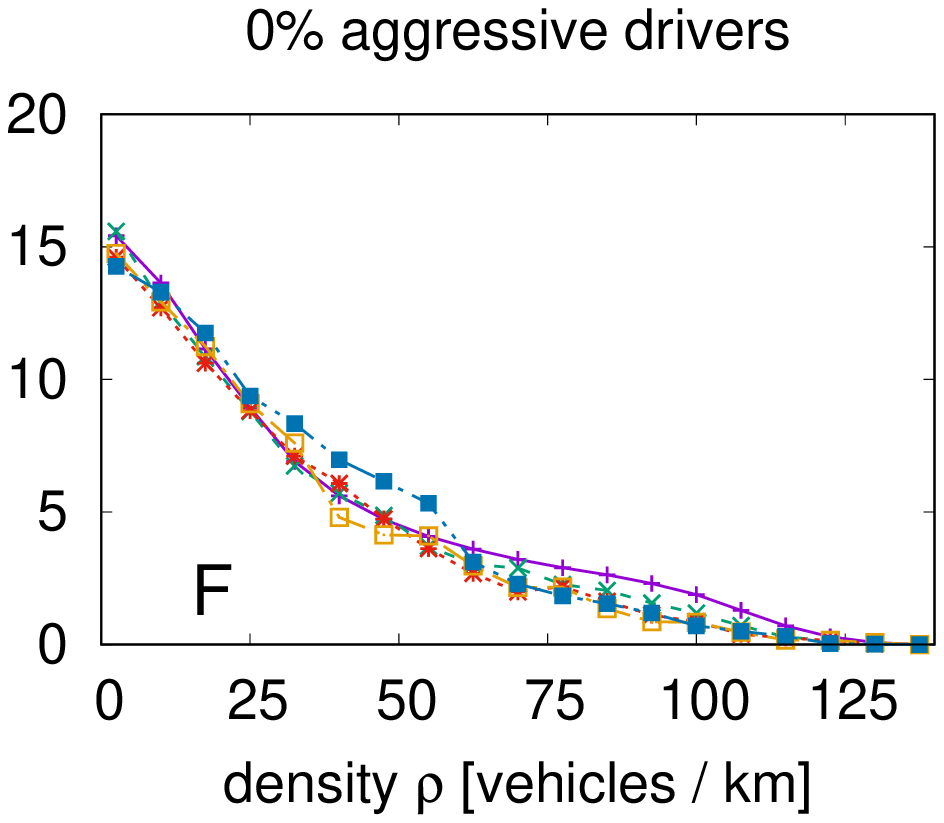}
\caption{Traffic flow (top row) and average velocity (bottom row) for the three configurations of aggressive and careful drivers using rand-mode traffic lights.}
\label{fig:rnd}
\end{figure}

\subsection{Gridlock}
Figure~\ref{fig:sync} shows traffic flow and average velocity for urban traffic using the three driving behavior configurations, the five turning probabilities and traffic lights in sync-mode. Figure~\ref{fig:rnd} shows similar results but setting traffic lights to rand-mode. The effect of turning is evident for sync-mode traffic lights. In rand-mode, turning probabilities make no significant differences for both the flow and the average velocity. Our results are similar to those reported in Ref.~\refcite{mhirech2011traffic} for urban traffic with signalized intersections.

The cases with 100\% aggressive drivers give higher fluxes at lower densities and traffic lights in sync-mode. This is no surprise, since sync-mode enables half of the city grid to momentarily behave as free roads during the green phase. However, the high flux is cut down abruptly to zero when a critical density is reached. This sharp change from a finite value to zero flow signals a total gridlock, meaning that all vehicles are standing still without moving. This constitutes a first order phase transition~\cite{chowdhury1999self,chowdhury2000statistical,kerner2011physics}. The cause of this lies in the distinctive feature of the aggressive type: the driver can block the intersection box. At low densities such behavior is of little or no consequence. But as soon as the streets become crowded and the drivers that are waiting to clear the intersection are actually blocking the street perpendicular to their motion, the traffic collapses and nobody can move.

Gridlocks emerge in sync- and rand-modes and for all turning probabilities whenever we have 100\% and 50\% aggressive drivers, as can be seen in panels A, B, D and E in Figs.~\ref{fig:sync} and \ref{fig:rnd}. The only cases in which gridlock is avoided is with 0\% aggressive drivers. The main feature of the careful driver is that she will never block the intersection.  She will wait until there is enough space ahead to actually traverse and clear the intersection. In cases of high vehicle densities, this behavior keeps the traffic flowing, as can be seen in panels C and F in Figs.~\ref{fig:sync} and \ref{fig:rnd}.

\begin{figure}
  \centering
\includegraphics[width=0.49\textwidth]{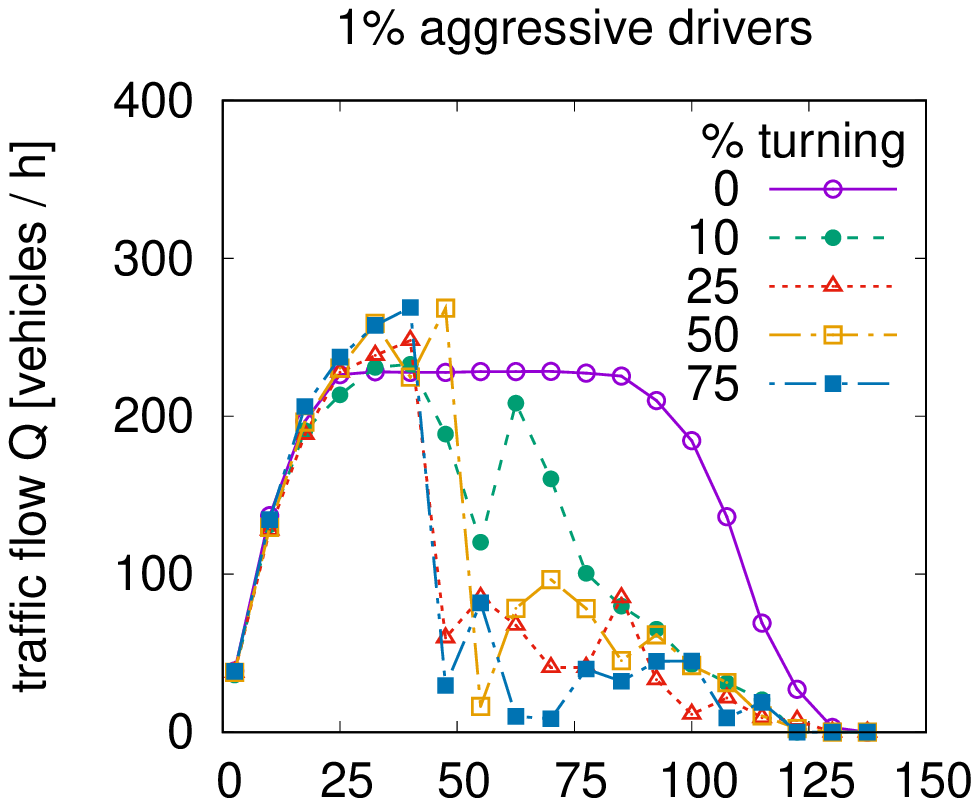}
\includegraphics[width=0.49\textwidth]{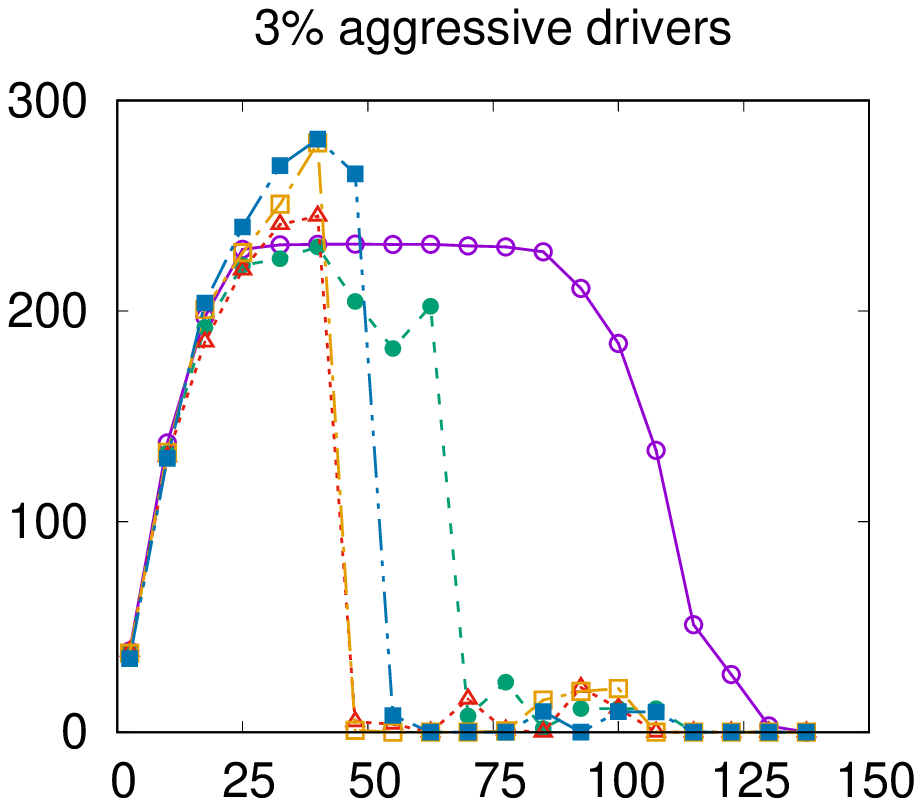}\\
\includegraphics[width=0.49\textwidth]{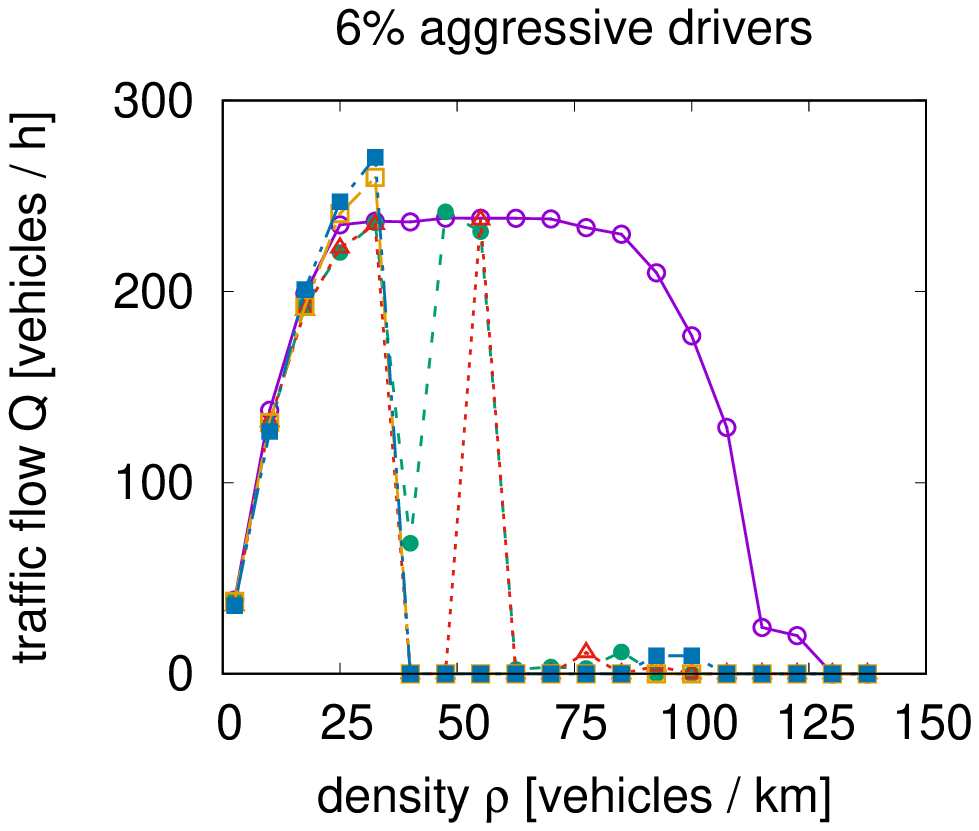}
\includegraphics[width=0.49\textwidth]{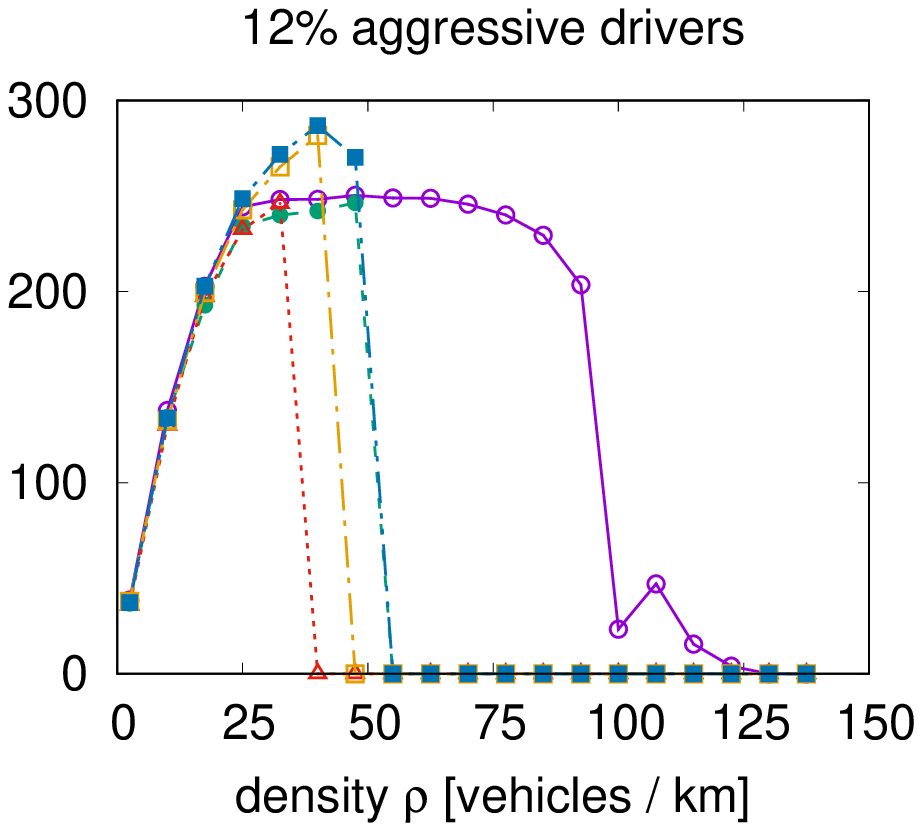}
\caption{Traffic flow for small fractions of aggressive drivers and rand-mode traffic lights. Gridlocks are formed with as little as 1-3\% aggressive drivers.}
\label{fig:min}
\end{figure}
There is always traffic flow in the 0\% aggressive configuration. It goes to zero only at high densities, when vehicles approach their allowed minimum separation. We do not find any traffic jam when all drivers are careful. This poses the question: how many aggressive drivers are needed in order to produce a gridlock? We explored this situation performing runs with 1, 3, 6, 12 and 25\% aggressive drivers, the traffic flow for these cases (25\% not shown) are presented in Fig.~\ref{fig:min}. We can see that traffic jams arise with as little as 1-3\% aggressive drivers. In the 1\% case, gridlocks are more frequent with increasing turning probability. In the 3\% case, only the 10\% probability turning curve is free from gridlocks until $\rho\approx60$~vehicles/km, for higher turning probabilities gridlocks are present.

The effect of a higher turning probability is translated into a decrease in flow for the sync-mode traffic lights setting (Fig.~\ref{fig:sync}). In rand-mode it has no noticeable effect on the flow (panels A and D in Fig.~\ref{fig:rnd}). It makes the gridlock onset appear at lower densities~\cite{chowdhury2000statistical} (panels B and E) and correlates with a smaller flow for 0\% aggressive drivers (panels C and F). This can be understood noticing that an aggressive driver that takes a turn into a congested lane, can end up blocking the intersection, instead of just going ahead had it not turned.

We chose to simulate a three-hour period of congested traffic, which is enough to collapse the flow even with a small number of aggressive drivers. The time it takes for a gridlock to appear depends on the vehicle density. The higher the density the sooner it emerges. We can say that a traffic jam could be avoided if higher densities dissolve quickly, instead of lasting for as long as three hours. Consequently, the system would have a higher tolerance to aggressive drivers, requiring larger amounts of them for a gridlock to happen.  We therefore take the three-hour period as a worst case scenario.

\section{Conclusion}
\label{sec:conclusion}
We have formulated a microscopic traffic model that simulates the decision making process for two kinds of driving behaviors. The formulation is based on the values of three parameters at a given instant of time. These parameters are the following distances: to the leading vehicle, to the next traffic light and to the nearest stopping point when the next traffic light is in the red phase. More complex decision trees can be formulated making use of additional parameters. This would account for a more realistic driving behavior. For instance, the driver could base her decision of going through the intersection not only on the location of her leading vehicle but also on the location of the next one ahead.

We have shown the importance of individual driving behavior in the collective properties of traffic flow, specifically in the phase transition from congested flow to a traffic jam or gridlock. In practice it is impossible to have the ideal situation of attaining zero aggressive drivers; however, the less they are, the better the chances to keep traffic flowing at rush hours.

Gridlocks arise even when very few aggressive drivers are present. Around 3\% is enough to collapse the traffic flow. This is a very important lesson in promoting good driving behavior, since very few aggressive drivers can create chaos, it is even more important that everybody do their best not to be one of them.

Rand-mode traffic lights give more similar fluxes and speeds regardless of turning frequency. This is because rand-mode does not enable a preferred direction of motion as sync-mode does by shifting all traffic lights at the same time. Randomization makes the flow insensitive to turning frequency, making gridlocks appear at a more defined value of density. We consider that rand-mode is the scheme that best represents the traffic light conditions in the portion of Guatemala City called {\it Centro Histórico}.

Several aspects of city traffic have been studied with cellular automata~\cite{schadschneider2002traffic, paissan2013imitation, fan2014characteristics, qi2016impact, szajowski2020drivers}. Our results support their findings in the sense that drivers who do not follow good conduct rules tend to worsen the traffic. Here we offer a different approach to account for driver behavior.

Future work includes the effects of the extent of the traffic light cycle and the use of different cycle lengths for different intersections.

\section*{Acknowledgments}
Motivation and inspiration for this paper came from the many hours lost in traffic during rush hours in Guatemala City.

\bibliographystyle{ws-ijmpc}
\bibliography{references}

\end{document}